\documentclass[11pt,titlepage]{article}
\usepackage{setspace}
\usepackage{epsfig}
\usepackage{ae}
\usepackage{aecompl}
\usepackage{amsmath}
\usepackage{graphicx}
\usepackage[round,numbers,sort&compress]{natbib}
\begin{document}

\title{How cholesterol could be drawn to the cytoplasmic leaf
  of the plasma membrane by phosphatidylethanolamine}

\author{
 H.  Giang and   M. Schick\\
Department of Physics, 
University of Washington,Seattle, WA 98195}

\date{\today} 
\maketitle 
\begin{abstract} 
In the mammalian plasma membrane, cholesterol can translocate rapidly 
between the exoplasmic and cytoplasmic leaves, so that its 
distribution between them should be given by the equality of its 
chemical potential in the leaves. Due to its favorable interaction
with sphingomyelin, which is almost entirely in the outer leaf, one
expects the great majority of cholesterol to be there also. Experimental
results do not support this,  implying that there is some mechanism
which attracts cholesterol to the inner leaf. 
We hypothesize that it is drawn there to reduce 
the bending free energy of the membrane caused by the presence of 
phosphatidylethanolamine (PE).  It does this in two ways: first by simply
diluting the amount of PE in the inner leaf, and second by ordering the
tails of the PE so as to reduce its spontaneous curvature. 
Incorporating this mechanism into a model free 
energy for the bilayer, we find that between 50 and 60\% of the 
total cholesterol should be in the inner leaf of human erythrocytes.
\end{abstract} 
\newpage 
\section{Introduction} The importance of cholesterol in the 
regulation of the properties of mammalian cells is widely recognized, 
and it has been the subject of intense research  \cite{yeagle85,maxfield10}. 
Nonetheless, some very basic questions about it remain 
unanswered. Among these is its distribution between the two leaves of the 
plasma membrane. It is well known that cholesterol can translocate rapidly 
between these leaves \cite{lange81,muller02,steck02}. As a consequence, 
its distribution should be determined from the equilibrium requirement 
that the chemical potential of cholesterol be the same in both leaves. 
Given this, the well-known preference of cholesterol for sphingomyelin 
(SM) among phospholipids \cite{niu02}, and the fact that almost all of the 
SM is in the exoplasmic leaf of the plasma membrane \cite{devaux91}, one 
might expect that the free energy of the system would decrease with an 
increase in the concentration of cholesterol in the outer leaf, and that 
the cholesterol would be found predominantly in that 
leaf. Indeed molecular dynamics simulations of some simple models of 
asymmetric bilayers incorporating SM and cholesterol do find the majority 
of cholesterol in the outer leaf \cite{perlmutter11,polley12}, just as
early experiments did \cite{fisher76}.

 But since 1982, experiments have consistently reported distributions of 
cholesterol which are contrary to 
these expectations. There is less agreement on what the actual distribution  
is. Some estimate that the cholesterol is rather evenly divided
between leaves \cite{lange82,muller02}, others that it is found 
to a greater extent in the inner, cytoplasmic, leaflet 
of the plasma membrane of various cells 
	\cite{brasaemle88, schroeder91,wood90,igbavboa96,mondal09}. 
Initial observations were made on 
human erythrocytes \cite{brasaemle88, schroeder91} with 75 to 80\% of the 
cholesterol reported to be in the inner leaf. Similar observations were 
then made in plasma membranes of other cells, such as neurons in mice 
\cite{wood90,igbavboa96}, ovaries in Chinese hamsters \cite{mondal09}, and 
the endocytotic recycling compartment of the latter \cite{mondal09}. 
Nonetheless, because the experimental evidence can be characterized as 
indirect, and a rationale for the results is absent, a recent review could 
describe the current situation by stating that ``... the transbilayer 
orientation of the sterols that make up one-third of the lipids in the 
eukaryotic plasma membrane has still not been resolved satisfactorily'' 
\cite{meer11}.

In this paper we propose two related mechanisms that would counteract
the attraction of cholesterol for the SM in the outer, exoplasmic, 
leaflet and would
draw it to the inner, cytoplasmic, one. We begin with the 
observation that almost all of the phospha\-tidyl\-ethanol\-amine (PE) is 
in the cytoplasmic leaf \cite{devaux91}. PE has a small head group, and 
thus a spontaneous curvature which is relatively large 
\cite{kollmitzer13}. Because of this, PE forms inverted hexagonal phases 
at high temperatures, at which the entropy of its hydrocarbon tails 
dominates, and forms lamellar phases only at lower temperatures 
\cite{tilcock82}. Thus the free energy of bilayers containing PE in the 
inner leaf must encompass a significant amount of bending energy. This 
bending energy is quadratic in the concentration of PE, and therefore acts 
like a repulsive interaction between PE molecules.  Such an interaction is 
equivalent to an attractive interaction between PE and all other 
components, and affects their distribution. In particular, the bending
energy can be reduced simply by
diluting the PE and replacing it with any
other component which does not increase the spontaneous curvature of the
leaf. We assume that this is true of cholesterol, due to its small
size and its placement below the head groups of the
phospholipids, as in the umbrella model \cite{huang99}.

In addition to this, the bending energy penalty 
is also quadratic in the spontaneous curvature of the PE, which increases
with the disorder of its tails. But
cholesterol is known to decrease the disorder of 
hydrocarbon tails of phospholipids \cite{hung07}. In particular, 
in excess of  0.35 mol fraction cholesterol, 
palmitoy\-loleoyl\-phosphatidyl\-ethanol\-amine (POPE)
bilayers are very well ordered, comparable to those of 
palmitoyloleoylphosphatidylcholine, (POPC),
cholesterol bilayers \cite{pare98}. Thus cholesterol will be 
drawn to the inner leaflet to reduce the bending energy penalty of PE by 
decreasing its spontaneous curvature. 
That this bending energy penalty 
is reduced by a sufficient concentration of cholesterol is in accord with 
the sterol's effect on the temperature of transition of PE from the 
high-temperature hexagonal phase to the low-temperature lamellar one. 
Whereas the initial addition of cholesterol decreases this 
transition temperature and stabilizes the inverted-hexagonal phase,  
amounts greater than 0.3 mol fraction {\em increase} the transition temperature 
and stabilize the lamellar phase \cite{epand87,takahashi96}. 
Some lipids, like palmitoyloleoylphosphatidylserine, (POPS), and
POPC, also stabilize a lamellar phase in mixtures with PE
\cite{epand88}, but do so because their own 
architecture and interactions favor a lamellar phase, and not because
they order the chains of PE. Thus, in contrast to cholesterol, they are
not expected to relieve the bending energy penalty of incorporating PE
into bilayers. 

In sum, we suggest that the free energy of the system is decreased if 
cholesterol is drawn to the cytoplasmic leaf of the plasma membrane 
because it reduces the bending 
energy caused by the inclusion of PE.  It does this in two ways: first, 
by simply diluting the PE in the inner leaf; second, by actually
reducing the magnitude of the spontaneous curvature of the PE itself.

We incorporate this hypothesis into a model of an asymmetric membrane 
consisting 
of phosphatidylcholine (PC), SM, and cholesterol in the outer leaf and 
phosphatidylserine (PS), PE, and cholesterol in the inner leaf. We take from 
experiment on erythrocytes
the ratios of SM to PC and of PE to PS  as well 
as the ratio of the total amount of cholesterol to the total amount of 
lipid. Requiring that the chemical potential of cholesterol in the two 
leaves be the same, we determine the fraction of cholesterol in the inner 
leaf. For reasonable values of the interaction parameters we find, in
the absence of
the bending energy penalty, that only about 25\% of the 
cholesterol is in the inner leaf. The majority is in the outer leaf 
due to the presence of SM there. 
Including the bending energy, but ignoring any effect of cholesterol on
the spontaneous curvature of PE, 
we determine that about 39\% of cholesterol is now in the inner
leaf. Finally, taking account the ordering effect of cholesterol on PE,
we find that
between 50 and 60\%  of the cholesterol should be found in the
cytoplasmic leaf of the human erythrocyte membrane.

\section{Theoretical Model}
\subsection{Procedure}
We consider a bilayer of which the outer, exoplasmic, leaf consists of
$N_{SM}$ molecules of sphingomyelin, $N_{PC}$ molecules of
phosphatidylcholine, and $N_{C_o}$ molecules of cholesterol, and the
inner, cytoplasmic leaf, consists of $N_{PE}$ molecules of 
phosphatidylethanolamine, $N_{PS}$ molecules of phosphatidylserine, and
$N_{C_i}$ molecules of cholesterol.  We assume that the glycerophospholipids
are, for the most part, unsaturated in the sn-2 chain. 
We denote the total number of
molecules in the outer leaf by $N_o$, the total number of molecules in
the inner leaf by $N_i$, and the total number of molecules in the
bilayer by $N_{bi}$. We assume that each leaf is a
liquid with the areas of the outer leaf, $A_o$, and of the
inner leaf, $A_i$,  directly related to their 
molecular compositions. If the area per molecule of the phospholipids be
denoted by $a$ and that of cholesterol by $r_aa$, then
\begin{eqnarray}
A_o&=&[N_{SM}+N_{PC}+r_aN_{C_o}]a=[N_o-(1-r_a)N_{C_o}]a\nonumber\\
\label{areas}
A_i&=&[N_{PE}+N_{PS}+r_aN_{C_i}]a=[N_i-(1-r_a)N_{C_i}]a
\end{eqnarray} 
In the absence of lateral pressure,
the Helmholtz free energy of the bilayer,
$F_{bi},$ depends only on the temperature, $T$, and the numbers of
molecules of each component. As the free energy is an extensive
quantity, it can be written in the form
\begin{eqnarray}
\label{free1}
&&F_{bi}(N_{SM},N_{PC},N_{C_o},N_{PE},N_{PS},N_{C_i},T)=\nonumber\\
&&\qquad N_{bi}f_{bi}(x_{SM},
x_{PC},x_{C_o},x_{PE},x_{PS},x_{C_i},T),
\end{eqnarray}
where $N_{bi}=N_o+N_i$ and $x_{SM}\equiv N_{SM}/N_{bi},$ is the mol
fraction of SM in the
bilayer and similarly for the other components. By definition
the sum of the mol fractions of all components is unity.

 It will be more convenient to express quantities in terms of the
 mol fractions of a component in the inner or outer leaf rather than the
 mol fraction in the bilayer. Thus we introduce 
\begin{eqnarray}
&&y_{SM}=\frac{N_{SM}}{N_o}=x_{SM}\frac{N_{bi}}{N_o},\qquad 
y_{PC}=x_{PC}\frac{N_{bi}}{N_o},\qquad y_{C_o}=x_{C_o}\frac{N_{bi}}{N_o},\\
&&y_{PE}=\frac{N_{PE}}{N_i}=x_{PE}\frac{N_{bi}}{N_i},\qquad
  y_{PS}=x_{PS}\frac{N_{bi}}{N_i},\qquad y_{C_i}=x_{C_i}\frac{N_{bi}}{N_i}.
\end{eqnarray}
By definition $y_{SM}+y_{PC}+y_{C_o}=1$ and $y_{PE}+y_{PS}+y_{C_i}=1$, 
so that only four of these mol fractions  are independent.
We take the areas of the two leaves to be equal as the fractional 
area difference, being of the order of the ratio of the membrane
thickness to the cell diameter, is small, of order $10^{-3}$. 
While this difference can be  
of importance for the gross morphology of cells \cite{wortis08}, 
it is not of importance here.
From the equality of areas,
the fractions $N_{bi}/N_i$ and $N_{bi}/N_o$ can be obtained, 
and the mol fractions $x$ can be expressed in terms of the mol fractions
$y$ and {\em vice versa}. In particular, the total mole fraction of
cholesterol in the bilayer, $x_C$, is given by
\begin{equation}
\label{totcholy}
x_C=\frac{y_{C_i}+y_{C_o}-2(1-r_a)y_{C_i}y_{C_o}}{2-(1-r_a)(y_{C_i}+y_{C_o})}.
\end{equation}
The four independent mol fractions, then, are determined by 
the requirement that the chemical
potentials of cholesterol in the two leaves be the same,
that the ratios of SM to PC in the outer leaf, $y_{SM}/y_{PC},$ and
of PS to PE in the inner leaf, $y_{PS}/y_{PE}$ in the inner leaf be
equal to their experimental values, and that
the total mol fraction of
cholesterol in the bilayer, Eq. (\ref{totcholy}), be equal to its
experimental value. Then the distribution of cholesterol between leaves
is determined.

We now turn to three models for the free energy of the bilayer.  In the 
first, we consider each leaf to be described by a phenomenological, 
regular solution, free energy \cite{furman77}.  The   
coupling between leaves is provided solely by the equality of the
cholesterol chemical potentials in the two leaves. We do this because it
is not only unclear what
other coupling mechanisms are important \cite{may09}, but also because
none of them so directly
affect the cholesterol distribution as the one we do incorporate.
In the second model, we  include a bending energy 
simply due to the presence of PE in the inner layer. This
draws 
the cholesterol to the cytoplasmic leaf where it reduces 
the bending 
energy penalty caused by the presence of PE simply by diluting it. 
Finally we include the effect on 
the spontaneous curvature of PE due to cholesterol, which  draws  additional  
amounts of it to the inner leaf to further reduce the bending energy caused by the 
presence of PE.

\subsection{Regular Solution Free Energy} 
We take as the model free energy a sum of the free energies of the 
two leaves in the form
\begin{eqnarray}
&&F_{bi}(N_{SM},N_{PE},N_{C_o},N_{PE},N_{PS},N_{C_i},T)=
N_of_o(y_{SM},y_{PC},y_{C_o},T)+N_if_i(y_{PE},y_{PS},y_{C_i},T),\nonumber\\
\label{finner}
&&f_i=6\epsilon_{PS,PE}y_{PS}y_{PE}+6\epsilon_{PS,C}y_{PS}y_{C_i}+
6\epsilon_{PE,C}y_{PE}y_{C_i}+\nonumber\\
&&\qquad k_BT(y_{PS}\ln y_{PS}+y_{PE}\ln y_{PE}+y_{C_i}\ln y_{C_i}),\\
\label{fouter}
&&f_o=6\epsilon_{SM,PC}y_{SM}y_{PC}+6\epsilon_{SM,C}y_{SM}y_{C_o}+
6\epsilon_{PC,C}y_{PC}y_{C_o}+\nonumber\\
&&\qquad k_BT(y_{SM}\ln y_{SM}+y_{PC}\ln y_{PC}+y_{C_o}\ln y_{C_o}).
\end{eqnarray}
We have assumed an average of six nearest-neighbor 
interactions per molecule. From this free energy we calculate the chemical 
potential of the cholesterol in the inner and outer leaves.

\begin{eqnarray}
\mu_{C_i}&=&\frac{\partial F_{bi}}{\partial N_{C_i}}=\frac{\partial 
N_if_i(y_{PE},y_{PS},y_{C_i},T)}{\partial N_{C_i}}\nonumber\\
         &=& \frac{\partial f_i}{\partial 
y_{C_i}}+f_i-\sum_j\frac{\partial f_i}{\partial y_j}y_j,\qquad j=PE,PS,C_i,\\
\mu_{C_o}&=&\frac{\partial F_{bi}}{\partial N_{C_o}}=\frac{\partial 
N_of_o(y_{SM},y_{PC},y_{C_o},T)}{\partial N_{C_o}}\nonumber\\
         &=& \frac{\partial f_o}{\partial y_{C_o}}+
       f_o-\sum_k\frac{\partial f_o}{\partial y_k}y_k,\qquad k=SM,PC,C_o.        
\end{eqnarray}
Again, to determine the six mol fractions, we equate these two chemical 
potentials, utilize the two constraints
\begin{eqnarray}
\sum_jy_j&=&1\qquad j=PE,PS, C_i,\\
\sum_ky_k&=&1\qquad k=SM,PC,C_o,
\end{eqnarray}
and set to their experimental values the ratios of SM to PC in the outer 
leaf, $y_{SM}/y_{PC}$, of PS to PE in the inner leaf, $y_{PS}/y_{PE}$, and 
the total mol fraction of cholesterol in the bilayer, $x_C$, Eq. 
(\ref{totcholy}). Once $y_{C_i}$ and $y_{C_o}$, the mol fractions 
of cholesterol in each leaf are obtained, the percent of the total cholesterol in the inner leaf 
follows from
\[p=\frac{y_{C_i}}{y_{C_i}+y_{C_o}(N_o/N_i)}\times 100,\]
where the ratio of the number of molecules in each leaf is determined by
the equality of the areas, Eq. \ref{areas}.

We must now set the parameters of our model. For the ratio of the area
per molecule of cholesterol to the area per molecule of the other
phospholipids, we take $r_a=0.6$ because the average area per molecule of
phospholipids is on the order of $a=0.7$ nm$^2$ and that of cholesterol is
about 0.4 nm$^2$ \cite{phillips72,hung07}. For the binary interactions,
we choose $\epsilon_{SM,C}=-0.58\,k_BT$, $\epsilon_{PC,C}=0.2\, k_BT$,
$\epsilon_{SM,PC}=0.30\ k_BT$,
$\epsilon_{PS,C}=-0.06\ k_BT$, $\epsilon_{PE,C}=0.28\ k_BT$,
and $\epsilon_{PS,PE}=0$. We discuss the
selection of these values in the Appendix.

We must also specify the membrane. We assume that it is at a temperature
$T=37^{\circ}C.$  We take the ratios of the components to be those of the human erythrocyte 
as given by Zachowski \cite{zachowski93}. 
There we find that the SM accounts for 0.22
of phospholipids in the outer leaf and 0.02 in the inner leaf, while PC
accounts for 0.20 of the phospholipids in the outer leaf and 0.07 in the
inner. For simplicity, we assume that all the SM and PC are in the outer
leaf and take the ratio $y_{SM}/y_{PC}=0.22/0.20=1.1.$ Similarly PS accounts
for 0.13 of phospholipids in the inner leaf and 0.02 in the outer, while
PE accounts for 0.25 in the inner leaf and 0.08 in the outer. Assuming
that all PS and PE are in the inner leaf, we take the ratio 
$y_{PS}/y_{PE}=0.13/0.25=0.52.$ Lastly we set the total mol fraction of 
cholesterol in the bilayer to be $x_C=0.4$ \cite{meer11}. 

It is now straightforward to carry out our program, and we find 
a single solution of our equations. By examining the matrix of second derivatives 
of the free energy, we have verified that this solution is stable. 
The inner leaf contains  a mol fraction of cholesterol
$y_{C_i}=0.22$. This corresponds to only 25\% of the total cholesterol 
being in the inner leaf.
This is easy to understand as
the energy of cholesterol is reduced if it goes to the
outer layer where it can interact favorably with the SM concentrated
there. It is easy to understand, but clearly not in accord with the reported results of
experiment. What physics is missing?

\subsection{Addition of the Bending Energy} 
It is our hypothesis that what is missing is that cholesterol can
ameliorate the bending energy cost of having PE in the cytoplasmic leaf
both by diluting the PE and also by increasing the order of its tails,
 reducing its curvature.

We incorporate this hypothesis into our model by adding to the free energy
of the flat bilayer a bending energy. Because absolute values of the  
spontaneous curvatures of 
SM, PC, and PS  are an order of magnitude smaller than that of PE 
\cite{bystrom03,fuller03,boulgaropoulos12, kollmitzer13}, we consider
the curvature only of the latter 
and write the bending
energy as
\begin{eqnarray}
F_{b}&=&\left(\frac{A_0+A_i}{2}\right)\frac{\kappa}{2}y_{PE}^2H^2_{PE}
\nonumber\\
   &=&\frac{1}{2}[N_i+N_o-(1-r_a)(N_{C_o}+N_{C_i})]f_b,\\
\label{fbend}
  f_b&=&\frac{1}{2}a\kappa y_{PE}^2H^2_{PE},
\end{eqnarray} with $\kappa$ the bending modulus. We shall take 
$\kappa=44 k_BT$ which is appropriate for red blood cells \cite{evans83}.

Our model free energy is now
\begin{equation}
F_{bi}=N_of_o+N_if_i+\frac{1}{2}[N_i+N_o-(1-r_a)(N_{C_o}+N_{C_i})]f_b, 
\end{equation}
with $f_i$ and $f_o$ given by Eqs. (\ref{finner}) and (\ref{fouter}) and
$f_b$ by Eq. (\ref{fbend}). 
We calculate the chemical potential of the cholesterol in the outer
and inner layers. We then set the areas of the two leaves,
Eqs. (\ref{areas}), to be equal after which we obtain
\begin{eqnarray}
\mu_{C_o}&=&f_o+\frac{\partial f_o}{\partial
  y_{C_o}}-\sum\frac{\partial f_o}{\partial
  y_j}y_j+\frac{1}{2}r_af_b,\qquad j=SM,PC,C_o,\\
\mu_{C_i}&=&f_i+\frac{\partial f_i}{\partial
  y_{C_i}}-\sum_k\frac{\partial f_i}{\partial y_k}y_k
+\frac{1}{2}r_af_b\nonumber\\
&&  +[1-(1-r_a)y_{C_i}]\left[\frac{\partial
    f_b}{\partial y_{C_i}}-\sum_k\frac{\partial f_b}{\partial
    y_k}y_k\right],\ k=PE,PS,C_i.
\end{eqnarray}

We must now specify the intrinsic curvature of PE. 
We first take its value to be that in the absence of cholesterol,
\begin{equation}
H_{PE}=H^0_{PE},
\end{equation}
one which has been measured to be $H^0_{PE}=-0.316$nm$^{-1}$ 
\cite{kollmitzer13}. Repeating our procedure, 
we now find a solution corresponding to a stable bilayer in which
the inner layer contains  a mol fraction $y_{C_i}=0.32.$ This corresponds
to 38.6\% of the total cholesterol now being found in the inner leaf.
As stated
above, the reason for this increase from the previous 25\% is simple. 
The bending energy penalty
due to the presence of the PEs is equivalent to a pair-wise repulsion 
between them, and thus an attraction between  them and all other
components. Cholesterol in the outer layer responds, decreasing the
free energy of the system by going to the inner layer and  
diluting the effect of the PE.

We now consider the additional effect of cholesterol acting 
on the tails of PE which,
for sufficient mol fraction, will order the tails of PE and further reduce the
bending energy cost of having PE in a planar bilayer. We incorporate
this effect by having the spontaneous curvature of PE depend upon the
cholesterol mol fraction. We choose 
\begin{equation}
\label{H0yci}
H_{PE}(y_{C_i})=H^0_{PE}-B\frac{y_{C_i}}{y_{min}}+\frac{B}{\lambda}\left(\frac{y_{C_i}}{y_{min}}\right)^{\lambda},
\end{equation}
with $B=0.05,$ $y_{min}=0.3$,
and $\lambda=8$. This
description is
dictated by the following considerations. First, from the behavior of
the temperature of transition of PE from the inverted-hexagonal to
lamellar phase \cite{epand87}, we know that the addition of cholesterol
initially stabilizes the former phase with respect to the latter. This
could result from the cholesterol decreasing the free energy of the
inverted-hexagonal phase, or increasing the free energy of the lamellar
phase, or a combination of the two. We choose the second of these
possibilities by having the spontaneous curvature of PE become more
negative with the initial addition of cholesterol. This leads to the
second term in Eq.(\ref{H0yci}) above.  

\begin{figure}[htbp]
\includegraphics[width=.5\columnwidth] {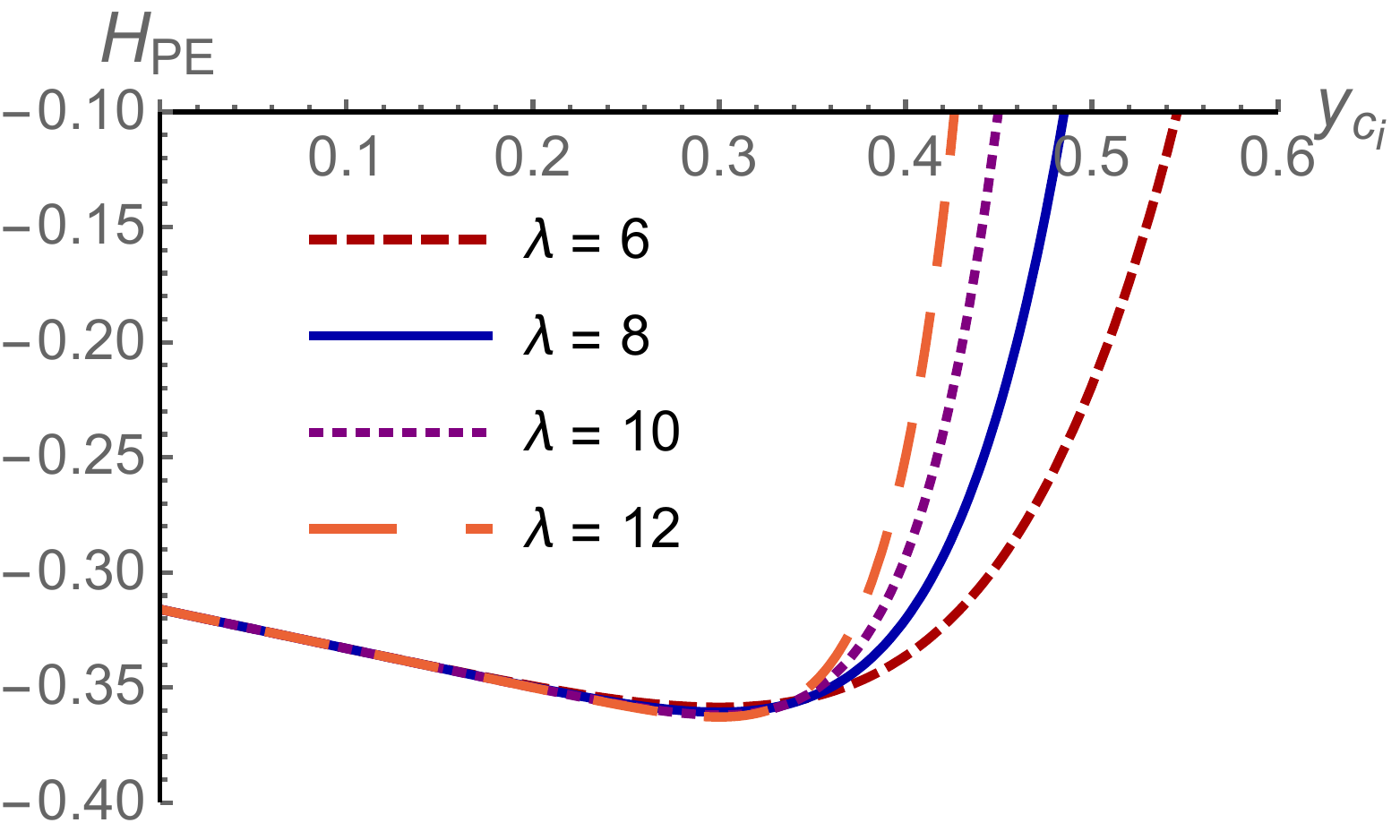}
\caption{Spontaneous curvature of PE in the presence of cholesterol 
as modeled by Eq. \ref{H0yci} for four values of $\lambda$} 
\label{H0}
\end{figure}

Second, and again from the behavior of the temperature of transition of
PE from the
inverted-hexagonal to lamellar phase \cite{epand87}, we expect that the 
magnitude of the spontaneous
curvature $H_{PE}(y_{C_i})$  no longer increases 
for mol fractions of cholesterol greater than $y_{min}\approx 0.3$. 
 Finally the
observation that, with a mol fraction cholesterol of 0.35,  
the tails of PE are as well-ordered as those of PC with
cholesterol \cite{pare98}, dictates a choice of $\lambda$ such that
the 
magnitude of the spontaneous
curvature of PE in the presence of cholesterol decreases rapidly for
values of $y$ somewhat larger than 0.35. The behavior of the spontaneous
curvature $H_{PE}(y)$ is shown in Fig.\ref{H0} for values of
$\lambda=6,\ 8,\ 10$ and 12. We choose $\lambda= 8.$ The dependence of our
results on our choice of $\lambda$  will be shown below.

With the above cholesterol-dependent PE spontaneous curvature, we now
find that the inner leaf contains $y_{C_i}=0.45$ mol fraction of cholesterol
 which corresponds to 58\% of the total cholesterol being in the inner leaf.
 We note that the mol fraction of cholesterol in the inner leaf is below the 
 maximum solubility of cholesterol in PE, 0.51 \cite{huang99}.
 With our solution, the mol fractions of the other components
in the plasma membrane are $y_{PS}=0.19$, $y_{PE}=0.36$ in the inner leaf
and $y_{C_o}=0.35$, $y_{SM}=0.34$, and $y_{PC}=0.31$ in the outer leaf.

\begin{figure}[htbp]
\includegraphics[width=.5\columnwidth] {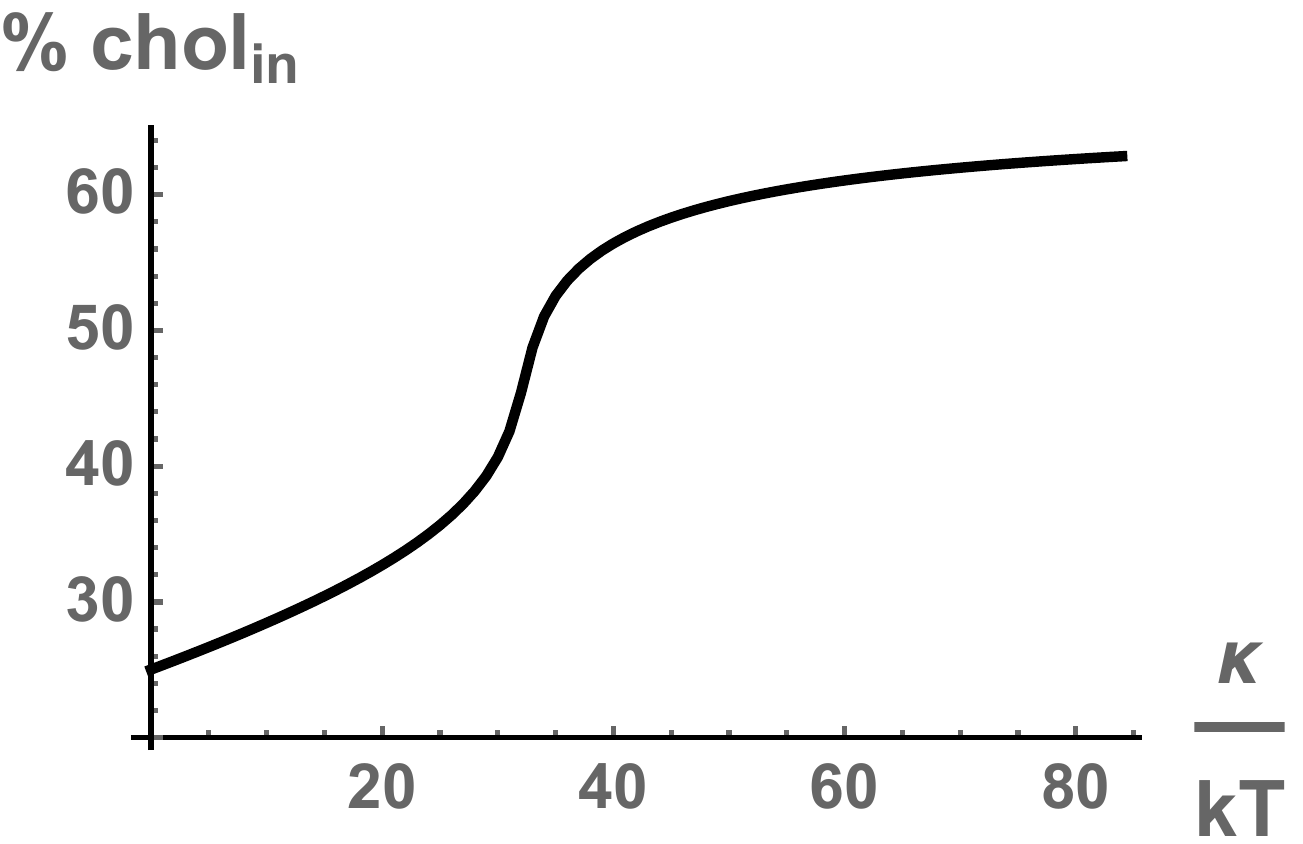}
\caption{Percent of cholesterol in the inner leaf as a function 
of the bending modulus $\kappa$.}
\label{cholkappa}
\end{figure}

As our hypothesis depends so directly upon the bending energy penalty,
our  results are clearly a function of the magnitude of the bending
modulus. 
We have taken it to be $\kappa=44k_BT$,  
appropriate for mammalian red blood cells \cite{evans83}.
The dependence
of the fraction of cholesterol in the inner leaf upon the bending
modulus is shown in Fig. \ref{cholkappa}.

\begin{figure}[htbp]
\includegraphics[width=.5\columnwidth] {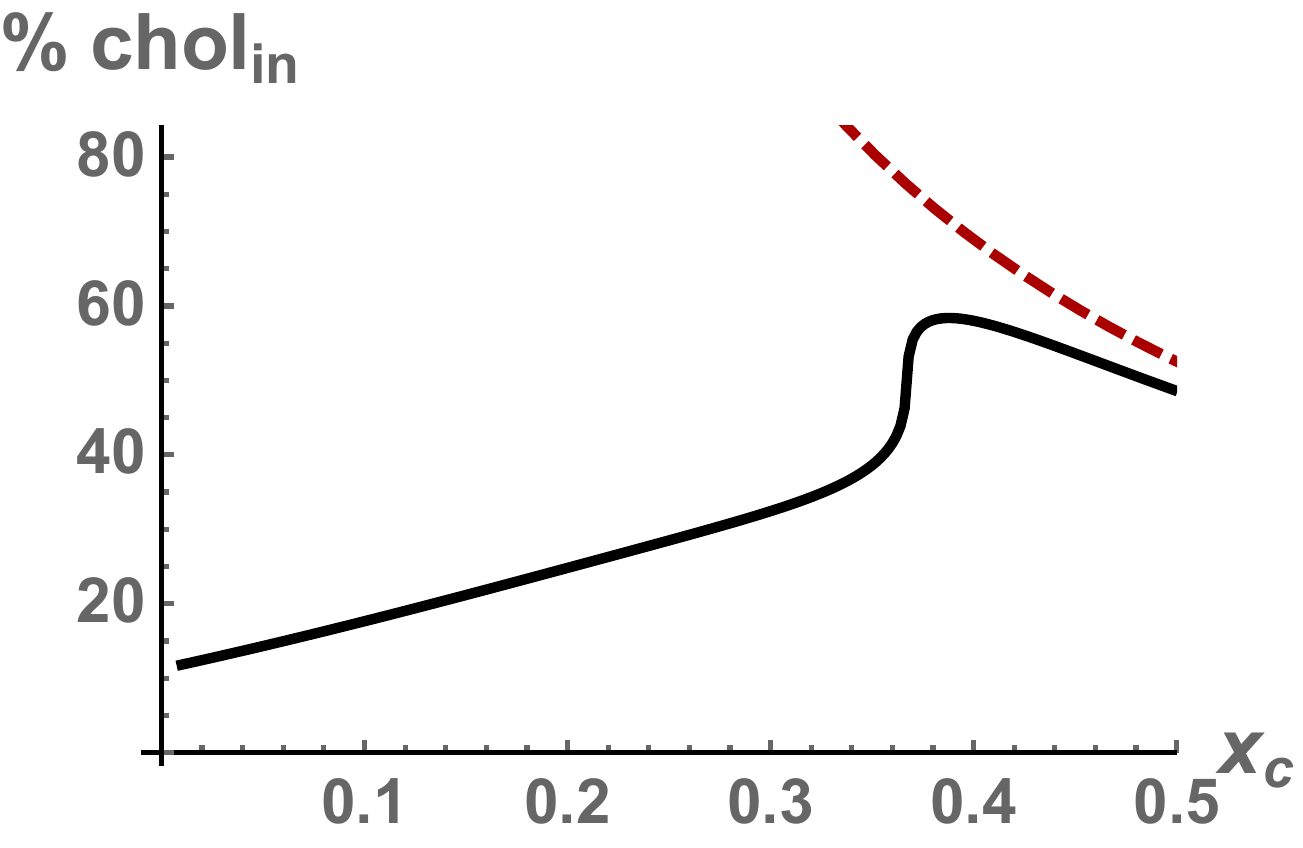}
\caption{Percent of total cholesterol in the inner leaf as a function 
of the total cholesterol mol fraction $x_C$. 
The dashed line shows the solubility limit 
of cholesterol.}
\label{cholxcA}
\end{figure}

We have taken the total cholesterol content of the membrane to be
$x_C=0.4$ appropriate for the plasma membrane of erythrocytes
\cite{meer11}. As for membranes of other cells, only the plasma membrane
and late-stage endosomes are characterized by such large levels of
cholesterol \cite{meer08}, while other membranes contain much less
cholesterol. It is of interest, therefore, to determine the dependence
of the percent of total cholesterol in the inner leaf as a function of
total cholesterol content. This is shown in Fig. \ref{cholxcA}. There
are a few things to note. First, when the total amount of cholesterol is
small, the percent in the inner leaf is also small. Hence 
for those membranes with small
amounts of cholesterol, we find that most of it will be in the outer
leaf. Second, the percent of cholesterol in the inner leaf initially
increases linearly. This is because of the reduction of the bending energy
penalty due to the dilution of the PE by cholesterol. The sharp
increase reflects the reduction by additional cholesterol 
of the PE spontaneous curvature.  Third,  
there is a maximum in the percent of cholesterol in the inner leaf as a
function of total cholesterol content. This follows from the fact that we
have found that the percent in the inner leaf can exceed 50\% and that
this percentage must approach 50 in the limit in which $x_C\rightarrow
1.$ It is interesting that the maximum occurs near $x_C\approx0.35$, not
very different from the value characterizing erythrocytes.

 \section{Discussion}

We have proposed that cholesterol is drawn to the cytoplasmic leaf of the 
plasma membrane in the plasma membrane because that is where the 
phospha\-tidyl\-etha\-nol\-amine is; that by going there, it reduces 
in two ways the
bending energy penalty of incorporating PE into a planar bilayer. 
First by simply diluting the PE mol fraction, cholesterol  
reduces the penalty which is quadratic in the PE mol fraction. Second,
in sufficient quantity, cholesterol orders the tails of PE thereby further
reducing the penalty of incorporating PE into the bilayer.
By these means, the chemical potential of cholesterol in the 
inner leaf will be negative, and matches the chemical potential of 
cholesterol in the outer leaf, which is negative due to the presence of 
sphingomyelin. We employed a simple model to show that in the absence of 
bending energy considerations, only about 25\% of the cholesterol would be 
in the cytoplasmic leaf,  again due to the presence of sphingomyelin in the 
exoplasmic leaf.  Including the bending energy penalty simply of
incorporating the PE, i.e. without any effect of cholesterol on its
tails, we found that the 
fraction of total cholesterol in the inner leaf would increase to 39\%.
Finally, by including the ordering effect of cholesterol on PE, we
obtained a fraction of total cholesterol in the inner leaf of some 58\%.

\begin{figure}[htbp]
\includegraphics[width=.5\columnwidth] {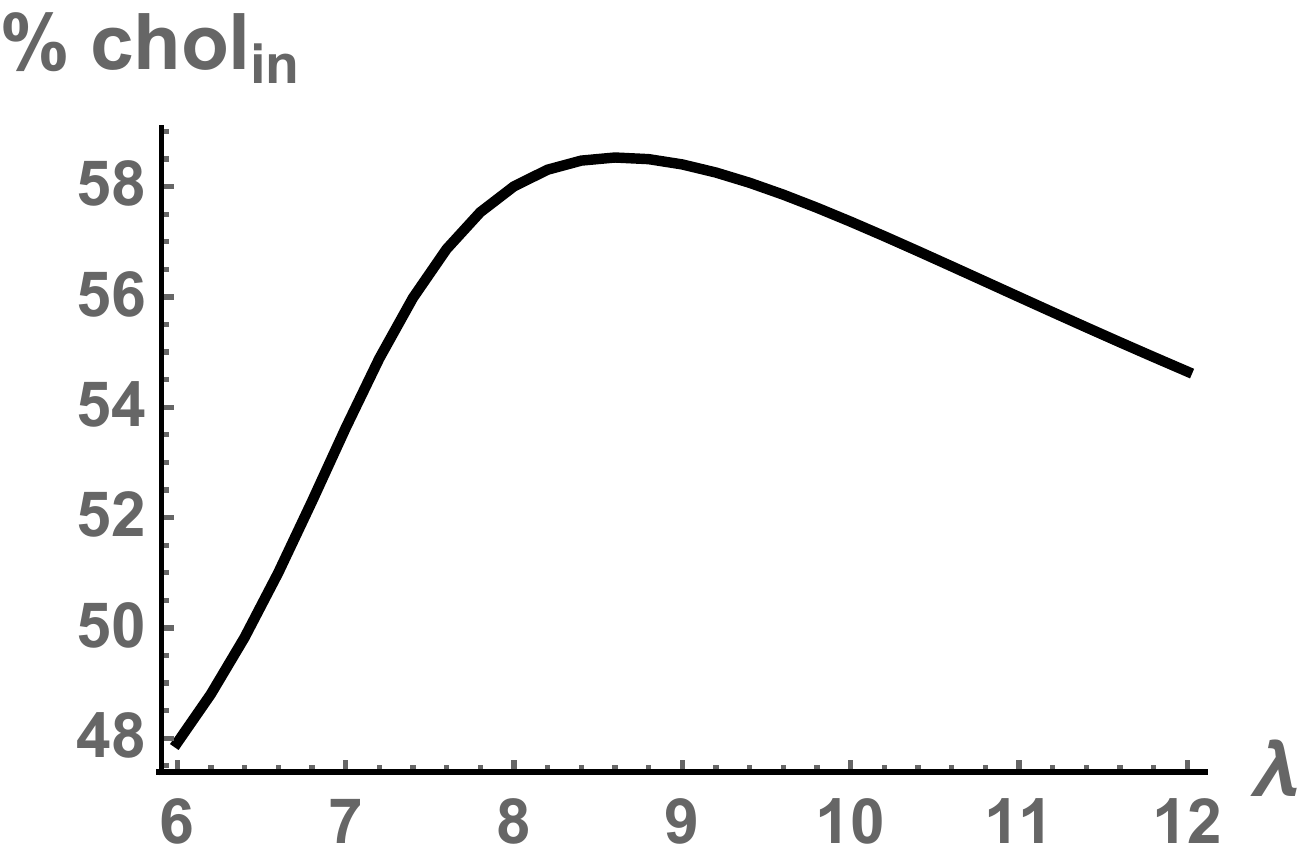}
\caption{Percent of cholesterol in the inner leaf as a function of the
  power $\lambda$
in the modeling of the spontaneous curvature, Eq. \ref{H0yci}.} 
\label{chollambda}
\end{figure}

Several comments are in order. The additional increase in the fraction
of cholesterol in the inner leaf due to the ordering of the PE tails by
cholesterol depends upon our phenomenological modeling of this effect
by our choice of the cholesterol-dependent spontaneous curvature of PE,
as given in Eq. \ref{H0yci}. The effect of varying the parameter $B$, which reflects 
the coupling of the cholesterol to the order of the PE tails is
simple. Were $B=0$, then the spontaneous curvature of PE would be
constant, equal to $H^0_{PE}$, and the only response of cholesterol
to the bending energy would be to dilute the PE. The fraction of the
total 
cholesterol which is in the inner leaf would be 39\%. This corresponds
to a mol fraction of cholesterol in the inner leaf of 0.32, too little
to order the tails of PE. 
For small values of $B$, this remains the case even though the
spontaneous curvature does depend on the amount of cholesterol in the
inner leaf. For values of $B$
greater than 0.045, however, the amount of cholesterol drawn to the inner
leaf is 0.45 mol fraction, sufficient to order the tails of the PE and
greatly reduce its curvature. The fraction of total cholesterol in the
inner leaf is 58\%, and is not very sensitive to further increases in
$B$.   
As to the parameter $\lambda$, we argued that a range of values
was permissible. We show in Fig. \ref{chollambda}
that the fraction of cholesterol in the inner leaf exceeds 50\% for the
values of $\lambda$ in this range.  In principle, one should be able to
go beyond this phenomenological modeling by utilizing more microscopic,
analytic, 
theories that include an explicit description of the lipid tails 
and can describe not only the polymorphism of PE \cite{li00a}, but also the
effect of cholesterol upon the tails \cite{elliott06}. Alternatively
simulations similar to, but larger than, those of Mori et al
\cite{mori04} should evince the effect we have described.   

 We have checked the dependence of our results on other assumptions
that we have made. The effect of reducing the number of nearest-neighbor
interactions from six to four is small; the percent of total cholesterol
in the inner leaf increases from 58 to 61\%.

 Because the cholesterol is
driven to the leaf in which the PE is located, the largest effect on our
result for the
cholesterol distribution comes
from including the amount of PE in the outer leaf, which is a fraction
0.08,
compared to 0.25 in the inner leaf \cite{zachowski93}. 
If we do this, as well as include the
0.07 PC in the inner leaf, compared to 
0.20 in the outer leaf, 
we find that for $x_C=0.4$ the percent of total cholesterol in the inner leaf
decreases from 58\% to 51\%.  

Even this modest majority, however, demonstrates our major point: that the
long-standing expectation that cholesterol should be found predominantly
in the
exoplasmic leaf of the plasma membrane because that is where the
sphingo\-myelin is located is very likely incorrect. The presence of
phospha\-tidyl\-ethanol\-amine in the cytoplasmic leaf at the cost of a
bending energy penalty attracts 
cholesterol there to reduce that penalty. The specific percent of
cholesterol in the inner leaf depends on several factors which we have 
tried to illustrate, but values on the order, or in excess, of 50\% can be understood.

We close with some observations concerning possible experiments.
First the effect of total cholesterol content on the percent of
cholesterol in the inner leaf, as shown in Fig. \ref{cholxcA}, should be
measurable in model asymmetric membranes which mimic the plasma 
membrane \cite{lin14}.

Second, as noted earlier, our model is consistent with results on the 
effect of cholesterol on the inverted-hexagonal to lamellar phase 
transition temperature in mixtures of cholesterol and POPE \cite{epand87}.  
There are some correlations clearly expected between cholesterol content 
and the specific form of PE in the membrane. Because the trans double bond 
in dielaidoylphosphatidylethanolamine is more easily ordered by 
cholesterol than is a cis double bond \cite{epand87,takahashi96}, one 
expects the asymmetry in the cholesterol distribution to be less.  It 
would be interesting to know how cholesterol affects a lipid with a 
polyunsaturated tail such as 18:0-20:4 PE as these polyunsaturated tails 
make up a non-negligible fraction, perhaps 18\%, of the chains of PE in 
the plasma membrane \cite{keenan70}. 

Finally, it is of interest to note that the effects of ergosterol on PE
membranes is not the same as the effect of cholesterol on them. While
the chain order of PE keeps increasing with cholesterol content to fractions on
the order of 0.45 mol fraction \cite{pare98}, the order saturates at ergosterol 
mol fractions of about 0.1 \cite{hsueh10}. Thus, as Richard Epand has
kindly pointed out to us, one might expect the fraction of
cholesterol in the inner leaf of yeast plasma membrane to be rather
different from that in mammalian cells. To our knowledge
the distribution of ergosterol between the leaves of the yeast
plasma membrane has not been measured. The result of such an experiment
would certainly be of interest and relevant to the considerations we
have put forth here.

\section{Appendix}
The free energy of a binary mixture of molecules of species $A$, and $B$, 
as obtained from regular solution theory is simply the mean-field 
approximation to the exact free energy obtained from a from a lattice-gas 
Hamiltonian of form

\begin{equation}
H=\sum_{<ij>}[E_{A,A}n^{A}_in^{A}_j+E_{B,B}n^B_in^B_j+
    E_{A,B}(n^A_in^B_j+n^B_in^A_j)]
\end{equation}
where $n^A_i=1$ if there is a molecule of species $A$ at the site $i$ and 
is zero otherwise, and similarly for $n^B_i$. The sum is over all distinct 
nearest-neighbor pairs of molecules. This Hamiltonian is easily mapped to 
that of an Ising model
\begin{equation}
H=-J\sum_{<ij>}S_iS_j-B\sum_iS_i,\qquad S_i=\pm1
\end{equation} 
via the relation $S_i=2n^A_i-1=1-2n^B_i$ so that the presence of an $A$ 
molecule is related to an up spin, and a $B$ molecule to a down spin. With 
this mapping, the interaction $J$ of the Ising model is then 
$J=\omega_{A,B}/2$ with 
$\omega_{A,B}\equiv{E_{A,B}-(E_{A,A}+E_{B,B}})/2.$ 
The exact 
transition temperature of the two-dimensional Ising on a triangular 
lattice is known to be $k_BT_c=4J/\ln 3=2\omega_{A,B}/\ln 3$ 
\cite{wannier45}. Thus if a physical system is known to undergo a phase
separation at a 
critical  
temperature $T$, then the interaction parameter in a model including 
fluctuations should be positive and taken to be 
$\omega_{A,B}=(1/2)\ln3k_BT\approx0.55k_BT.$ 

Regular solution theory, however, does not include fluctuations. 
Within this theory, the free energy per particle of a binary mixture 
on a triangular lattice can be written
\[f=6\epsilon_{A,B}y_Ay_B+k_BT(y_A\ln y_A+y_B\ln y_B),\]
It yields a transition  
$k_BT_c^{rs}=3\epsilon_{A,B}$. Therefore to obtain 
in regular solution theory the same transition temperature as the exact
result one must choose $\epsilon_{A,B}=(2/3\ln 3)\omega_{A,B}\approx
0.6\omega_{A,B}.$

The values of the interaction 
parameters, $\omega_{A,B},$ for many pairs of lipids can be estimated from 
experiment and 
have been collected by Almeida \cite{almeida09}. 
In particular for the interactions between components of the outer leaf 
at $T=37^{\circ}$C, $\omega_{PC,C}=0.34k_BT$, $\omega_{SM,C}=-0.97 k_BT,$ and 
$\omega_{SM,PC}=0.51k_BT$. As $\omega_{PC,C}$ is positive, we use in our
regular solution theory the estimate $\epsilon_{PC,C}/k_BT=
0.6\omega_{PC,C}/k_BT=  
0.20$. If we take the ratios of the other interactions parameters to be
the same as in the table, {\em i.e.} 
$\epsilon_{SM,C}/\epsilon_{PC,C}=\omega_{SM,C}/\omega_{PC,C},$ 
then $\epsilon_{SM,C}/k_BT=-0.58$ and similarly
$\epsilon_{SM,PC}/k_BT=0.30.$  

The interactions relevant to the inner leaf, those between PE and
cholesterol, between PS and cholesterol, and between PE and PS are not 
included in the table of Almeida. To obtain an estimate for them, we
proceed 
as follows: Cholesterol and PE do not phase separate at $T=37^{\circ}$
so that we should take $\epsilon_{PE,C}/k_BT<1/3$.  We choose 
$\epsilon_{PE,C}/k_BT=0.28.$ Next we estimate the interaction between 
PS and cholesterol. Niu 
and Litman \cite{niu02} have measured the differences 
$\Delta_{SM}\equiv\omega_{SM,C}-\omega_{PC,C}=-1181$ cal/mol=$-1.92k_BT$ 
at 37$^\circ$C, and $\Delta_{PS}=\omega_{PS,C}-\omega_{PC,C}=-0.65 k_BT.$
We assume that
 \[ \frac{\epsilon_{PS,C}-\epsilon_{PC,C}}{\epsilon_{SM,C}-\epsilon_{PC,C}}
 =\frac{\omega_{PS,C}-\omega_{PC,C}}{\omega_{SM,C}-\omega_{PC,C}}=\frac{0.65}{1.92}=0.34,\]
 so that $\epsilon_{PS,C}/k_BT=\epsilon_{PC,C}+0.34(\epsilon_{SM,C}-\epsilon_{PC,C})=-0.06.$
Lastly because the tails of PE and PS lipids are similar, 
we take $\epsilon_{PE,PS}/k_BT=0.$
This completes the specification of the interactions.

\section{Acknowledgments} We thank Paulo Almeida, Fred Maxfield, Alex Merz,
and Richard Epand in particular for 
useful correspondence and stimulating conversations. One of us, (MS), 
would like to thank William Clay 
and Josh Zimmerberg for their insights. This work is supported 
in part by the National Science Foundation under grant No. DMR-1203282.

 \clearpage
\bibliography{lipids14}

\end{document}